\documentclass[prd,12pt]{article}

\usepackage{tikz}
\usepackage{amsmath,amssymb,graphicx,multirow,xspace,slashed,array,booktabs}
\usepackage[colorlinks=true,urlcolor=blue,anchorcolor=blue,citecolor=blue,filecolor=blue,linkcolor=blue,menucolor=blue,pagecolor=blue]{hyperref}
\usepackage[compress,numbers]{natbib}
\usepackage{subfigure, placeins}
\usepackage{booktabs}
\usepackage{blindtext, rotating}
\usepackage{afterpage}
\usepackage{enumitem}
\usepackage{ marvosym }
\usepackage{authblk} 
\usepackage{verbatim}
\usepackage{soul} %scratch out text
\usepackage[normalem]{ulem}
\usepackage{pifont}% http://ctan.org/pkg/pifont

\usepackage[utf8]{inputenc}

\usepackage{floatrow}
\newfloatcommand{capbtabbox}{table}[][\FBwidth]

\usepackage[font=footnotesize,labelfont=bf]{caption}

\usepackage{lineno}

\allowdisplaybreaks

\long\def\symbolfootnote[#1]#2{\begingroup%
\def\thefootnote{\fnsymbol{footnote}}\footnote[#1]{#2}\endgroup}

\newcommand{\newc}{\newcommand}
\newc{\gsim}{\lower.7ex\hbox{$\;\stackrel{\textstyle>}{\sim}\;$}}
\newc{\lsim}{\lower.7ex\hbox{$\;\stackrel{\textstyle<}{\sim}\;$}}
\newc{\gev}{\,{\rm GeV}}
\newc{\mev}{\,{\rm MeV}}
\newc{\ev}{\,{\rm eV}}
\newc{\kev}{\,{\rm keV}}
\newc{\tev}{\,{\rm TeV}}
\newc{\MHT}{$H_T^{\text{miss}}$}
\newc{\MET}{$\slashed{E}_T$}
\newc{\MTT}{$M_{T2}$}

\newc{\mz}{M_Z}
\newc{\mpl}{M_*}
\newc{\mw}{m_{\rm weak}}
\newc{\nr}[1]{N^c_R{}_{#1}}

\definecolor{darkgreen}{rgb}{0,0.5,0}
\definecolor{goodorange}{rgb}{0.9,0.4,0}

\def\beq{\begin{equation}}
\def\eeq{\end{equation}}
\newcommand{\bea}{\begin{eqnarray}\begin{aligned}}
\newcommand{\eea}{\end{aligned}\end{eqnarray}}
\def\bitem{\begin{itemize}}
\def\eitem{\end{itemize}}

\numberwithin{equation}{section}

\newcommand\fverb{\setbox\fverbbox=\hbox\bgroup\verb}

\newbox\fverbbox

\newcommand\FLDs{F^L_{D^*}}

\newcommand\RD{{R_D}}
\newcommand\RDs{{R_{D^{*}}}}
\newcommand\RDRDs{{R_{D^{(*)}}}}
\newcommand\Rjpsi{{R_{J/\psi}}}
\newcommand\Bc{{{\rm Br}(B_c\to\tau\nu)}}
\newcommand\CO{{\mathcal O}}

\newcommand\CSRL{{C^S_{RL}}}
\newcommand\CSLL{{C^S_{LL}}}
\newcommand\CVLL{{C^V_{LL}}}
\newcommand\CVRL{{C^V_{RL}}}
\newcommand\CTLL{{C^T_{LL}}}

\newcommand\CSmL{{C^S_{-L}}}
\newcommand\CSpL{{C^S_{+L}}}
\newcommand\CVmL{{C^V_{-L}}}
\newcommand\CVpL{{C^V_{+L}}}

\begin{document}

\baselineskip 0.6cm

\begin{titlepage}

\thispagestyle{empty}

\begin{center}

\vskip 1cm

{\Huge \bf Maximizing the Impact of  }\vskip0.5cm
{\Huge\bf New Physics in $b\rightarrow c \tau \nu$ Anomalies}

\vskip 1cm

\vskip 1.0cm
{\large Pouya Asadi and David Shih }
\vskip 1.0cm
{\it  NHETC, Dept.~of Physics and Astronomy\\ Rutgers, The State University of NJ \\ Piscataway, NJ 08854 USA} \\
\vskip 1.0cm

\end{center}

\vskip 1cm

\begin{abstract}
We develop a rigorous, semi-analytical method for maximizing any $b\to c\tau\nu$ observable in the full 20-real-dimensional parameter space of the dimension 6 effective Hamiltonian, given some fixed values of $\RDRDs$. We apply our method to find the maximum allowed values of $\FLDs$ and $\Rjpsi$, two observables which have both come out higher than their SM predictions in recent measurements by the Belle and LHCb collaborations. 
While the measurements still have large error bars, they add to the existing $\RDRDs$ anomaly, and it is worthwhile to consider NP explanations. It has been shown that none of the existing, minimal models in the literature can explain the observed values of $\FLDs$ and $\Rjpsi$. Using our method, we will generalize beyond the minimal models and show that there is no combination of dimension 6 Wilson operators that can come within $1\sigma$ of the observed $\Rjpsi$ value. By contrast, we will show that the observed value of $\FLDs$ can be achieved, but only with sizable contributions from tensor and mixed-chirality vector Wilson coefficients.

\end{abstract}

\flushbottom

\end{titlepage}

\section{Introduction and summary}

Hints of new physics (NP) violating lepton flavor universality (LFU) have been observed in semileptonic $b$ decays, captured in the ratios   \cite{Aubert:2007dsa, Bozek:2010xy,Lees:2012xj, Lees:2013uzd, Aaij:2015yra,Huschle:2015rga, Abdesselam:2016xqt}
\begin{equation}
R_{D^{(*)}} = \frac{\Gamma (\bar{B} \rightarrow D^{(*)} \tau \nu)}{\Gamma (\bar{B} \rightarrow D^{(*)}  \ell \nu)},
\label{eq:rddef}
\end{equation}
where $\ell$ stands for either electrons or muons. The global average of the observed values is \cite{HFLAV16}  
\begin{equation}
R_D = 0.407 \pm 0.046 , \hspace{0.5 in}  R_{D^{*}} = 0.304 \pm 0.015,
\label{eq:rdobs}
\end{equation}
while the Standard Model (SM) prediction for these ratios is \cite{Lees:2012xj,Lees:2013uzd,Kamenik:2008tj,Fajfer:2012vx,Bailey:2012jg,Lattice:2015rga,Na:2015kha,Aoki:2016frl,HFLAV16,Jaiswal:2017rve} 
\begin{equation}
R_D^{SM} =0.299 \pm 0.003 , \hspace{0.5 in}  R_{D^{*}}^{SM} = 0.258 \pm 0.005.
\label{eq:rdsm}
\end{equation}  
This corresponds to a $\sim 3.8\sigma$ discrepancy with the Standard Model prediction \cite{HFLAV16}.\footnote{In this work we are not including the most recent Belle analysis on $\RDRDs$ \cite{Abdesselam:2019dgh}.}

A similar upward fluctuation has been observed in the following ratio as well
\begin{equation}
\Rjpsi = \frac{\Gamma (B_c \rightarrow J/\psi \tau \nu)}{\Gamma (B_c \rightarrow J/\psi \ell \nu)}.
\label{eq:rjpsidef}
\end{equation}
The value measured by LHCb is \cite{Aaij:2017tyk}
\begin{equation}
\Rjpsi =0.71 \pm 0.17\,\,({\rm stat}) \pm 0.18 \,\,({\rm sys}).
\label{eq:rjpsiexp}
\end{equation} 
There is significant uncertainty in the SM predictions for this ratio \cite{Dutta:2017xmj,Watanabe:2017mip,Tran:2018kuv,Cohen:2018dgz,Leljak:2019eyw}
\begin{equation}
R_{J/\psi}^{SM} \in \left( 0.2, 0.39 \right).
\label{eq:rjpsism}
\end{equation}

There are also a host of different polarization and asymmetry observables \cite{Tanaka:1994ay,Tanaka:2010se,Fajfer:2012vx,Sakaki:2012ft,Datta:2012qk,Duraisamy:2013kcw,Ivanov:2015tru,Becirevic:2016hea,Alonso:2016gym,Alok:2016qyh,Bardhan:2016uhr,Ivanov:2017mrj,Alonso:2017ktd,Asadi:2018sym} that can be measured in these decays. 
Recently, Belle has released preliminary results on the measurement of the $D^*$ longitudinal polarization fraction in the $B\rightarrow D^* \tau \nu$ decay \cite{Abdesselam:2019wbt}
\beq
\label{eq:fldexp}
\FLDs = 0.60 \pm 0.08\,\,({\rm stat})\pm 0.035\,\,({\rm sys}),
\eeq
where 
\beq
\label{eq:flddef}
\FLDs = \frac{\Gamma (\bar{B} \rightarrow D^*_L \tau \nu)}{\Gamma (\bar{B} \rightarrow D^* \tau \nu)}
\eeq
with $D_L^*$ referring to a longitudinally polarized $D^*$. Meanwhile the SM prediction is
\cite{Alok:2016qyh,Bhattacharya:2018kig,PhysRevD.98.095018}, e.g. \cite{Bhattacharya:2018kig} 
\beq
\label{eq:fldsm}
(F^{L}_{D^*})^{SM} = 0.457 \pm 0.010.
\eeq

While these seem to be interesting additions to the $\RDRDs$ anomaly, they are in tension with not only the SM prediction, but also various new physics models that have been considered in the literature \cite{Watanabe:2017mip,Tran:2018kuv,Leljak:2019eyw,Iguro:2018vqb,Blanke:2018yud}.\footnote{Ref.~\cite{Dutta:2017wpq} considers the possibility of right-handed (RH) neutrinos as well and reports pairs of WCs that are claimed to explain the observed $\Rjpsi$. We were unable to reproduce their results in our calculations.} In fact, no model has been found to come even close to the observed values of $\FLDs$ or $\Rjpsi$.

So far, only minimal BSM models (single mediators) and simple combinations of Wilson coefficients (WCs) have been considered. In this work, we will generalize the study of these observables to the full space of WCs for the dimension 6 effective Hamiltonian:
\beq\label{Heff}
 {\mathcal H}_{\rm eff} = \frac{4 G_F V_{cb}}{\sqrt{2}}   \sum_{X=S,V,T\atop M,N=L,R} C^X_{MN}{\mathcal O}^X_{MN} ,
\eeq
where the only WC generated in the SM is $\CVLL=1$, and the four-fermion effective operators are defined as
\begin{eqnarray}
 {\mathcal O}^S_{MN} & \equiv & (\bar c P_M b)(\bar \tau P_N \nu), \nonumber \\
{\mathcal O}^V_{MN} & \equiv &(\bar c \gamma^\mu P_M b)(\bar \tau \gamma_\mu P_N \nu), \label{eq:Lop}\\
{\mathcal O}^T_{MN} &\equiv &(\bar c \sigma^{\mu\nu} P_M b)(\bar \tau \sigma_{\mu\nu}P_N \nu), \nonumber
\end{eqnarray}
for $M,N = R$ or $L$. The two tensor operators $\mathcal{O}^T_{RL}$ and $\mathcal{O}^T_{LR}$ are identically zero; thus, the Hamiltonian includes 5 operators with either types of neutrinos. For simplicity, we will focus on operators with left-handed (LH) neutrinos in this work; 
then the full space of WCs consists of
\beq\label{Wilsonspace}
(C^V_{LL},C^V_{RL},C^S_{LL},C^S_{RL},C^T_{LL})
\eeq
which is 10 real dimensional. However, at the end of section \ref{sec:setup}, we will explain how our results can be straightforwardly generalized to the case of LH+RH neutrinos, leaving our conclusions unchanged.

Since the experimental error bars on $\FLDs$ and $\Rjpsi$ are much larger than those of $\RD$ and $\RDs$, it makes sense to treat the latter as constraints and attempt to maximize the former subject to those constraints. We will develop a fully general, rigorous, semi-analytical method to maximize essentially any $b\to c\tau\nu$ observable for fixed values of $\RD$ and $\RDs$. We will also fix ${\rm Br}(B_c\to\tau\nu)$ consistent with its upper bounds \cite{Li:2016vvp,Alonso:2016oyd,Celis:2016azn,Akeroyd:2017mhr}, as this was shown to play an important role in restricting the possible values of $\Rjpsi$ and $\FLDs$ \cite{Watanabe:2017mip,Iguro:2018vqb,Blanke:2018yud}. 

Using this approach, we find that the global maxima of $\FLDs$ and $\Rjpsi$, with $\RD$ and $\RDs$ fixed to their current world averages, and $\Bc\le 30\%$ are:
\beq
\FLDs \le 0.66,\qquad \Rjpsi\le 0.41
\eeq
We also explore values of $\RDRDs$ within their current 1 and 2$\sigma$ error ellipses, and different values of the $\Bc$ constraint. Our conclusions are qualitatively unchanged. 

We find that to reach the global maxima of $\FLDs$ and $\Rjpsi$, NP should give rise to the WCs $\CVRL$ and $\CTLL$ (or their  counterparts with RH neutrinos) and should partially cancel the SM contribution to $\CVLL$. (Intringuingly, the global maxima of $\FLDs$ and $\Rjpsi$ are characterized by very similar values of the WCs.) We will also show on completely general grounds that the observables are maximized for real-valued Wilson coefficients (up to an overall rephasing invariance).

Clearly, the observed value of $\Rjpsi$ cannot be explained with any combination of the dimension 6 Wilson operators.  If the current value of $\Rjpsi$ persists in future measurements (with reduced error bars), it will signify a major contradiction with the current framework. Either the numerical formula needs substantial revision (e.g.\ the hadronic form factors), or NP contributes in a way beyond the dimension 6 effective Hamiltonian (e.g.\ with very light mediators).

Meanwhile, we see that the current measured value of $\FLDs$ can be attained. To understand the ingredients necessary to reaching the current measured value, we further maximize $\FLDs$  with each WC held fixed. We will confirm using this approach that sizable $C^V_{RL}$ and $C^T_{LL}$ are required to come within $1\sigma$ of the current measured value of $\FLDs$, together with a modest amount of cancellation in $C^V_{LL}$. 

The need for $C^V_{RL}$ (or its RH neutrino counterpart) to account for $\FLDs$ is especially intriguing. It is well-known that these mixed-chirality vector operators are especially difficult to generate from any UV model, see \cite{Cerri:2018ypt} for a recent discussion and original references. Because they violate $SU(2)_L\times U(1)_Y$, they are higher effective dimension (requiring additional Higgs insertions), and so are generally absent or suppressed in any UV completion. Searching for a model that generates $C^V_{RL}$ or $C^V_{LR}$ is especially well-motivated now given our results.

Another reason previous studies may have failed to reach the measured value of $\FLDs$ is that we find multiple Wilson coefficients are necessary. This may point at nonminimal models, e.g.\ involving multiple leptoquarks.

As we have already noted, the experimental uncertainties on $\FLDs$ and $\Rjpsi$ (and the theoretical uncertainties on $\Rjpsi$) are still quite large, so the discrepancies in these observables may just be due to random fluctuations, and any attempt to read too much into them may be premature.
Nevertheless we feel a closer examination of these two observables is a useful exercise to attempt now, in that it may inspire interesting new directions in model building. The general method we develop for maximizing observables given the constraints, taming the huge parameter space of Wilson coefficients, may be of use to others interested in other observables, e.g.\ $R_{\Lambda_b}$. Finally, the study done here is something to keep in mind for the near future, 
where much more precise measurements of these observables with much more data from LHCb and Belle II are expected.

The outline of the paper is as follows. In Sec.~\ref{sec:setup} we explain our general approach for studying the space of all WCs. In Sec.~\ref{sec:max}, we will describe our results for the global maxima of $\FLDs$ and $\Rjpsi$ subject to the constraints.
In Sec.~\ref{sec:resultfixed} we  maximize the observables while fixing some of the WCs.

\section{General setup}
\label{sec:setup}

The observables of interest in this work are $\CO=\Rjpsi,\FLDs,\RD,\RDs,\Bc$. The first four observables show discrepancies with the SM predictions, while the bounds on $\Bc$ can be used to severely constrain various BSM explanations of these anomalies \cite{Li:2016vvp,Alonso:2016oyd,Celis:2016azn,Akeroyd:2017mhr}. Measurements of the total width of the $B_c$ meson and $B_u \rightarrow \tau\nu$ decay have been used in \cite{Li:2016vvp,Alonso:2016oyd,Celis:2016azn} and \cite{Akeroyd:2017mhr} to put bounds of $\Bc\lesssim 30\%$ and $\Bc\lesssim 10\%$, respectively. Meanwhile the SM prediction is $\Bc=2.3\%$. We will use these three reference values for $\Bc$ throughout this work. 

In our study of these observables, we use the numerical formulas in \cite{Iguro:2018vqb}, 
\bea
\label{eq:NformulasLHCV}
&\RD=0.299\left(|\CVpL |^2+1.02|\CSpL|^2+0.9|\CTLL|^2+{\rm Re}\left[(\CVpL)(1.49(\CSpL)^*+1.14(\CTLL)^*)\right]\right),\\
& \RDs =0.257\Big(0.95|\CVmL|^2+0.05|\CVpL|^2 + 0.04|\CSmL|^2+16.07|\CTLL|^2\\
&\qquad \qquad \qquad + {\rm Re}\left[\CVmL(0.11(\CSmL)^*-5.89(\CTLL)^*) \right]+0.77{\rm Re}\left[\CVpL (\CTLL)^*\right]\Big),\\
& \RDs \FLDs = 0.116\left(|\CVmL|^2+0.08|\CSmL|^2+7.02|\CTLL|^2+{\rm Re}\left[(\CVmL)(0.24(\CSmL)^*-4.37(\CTLL)^*)\right]\right),\\
&{\rm Br}(B_c\to\tau\nu) = 0.023 \left( |\CVmL + 4.33 \CSmL|^2 \right),
\eea
where we are defining $C^S_{\pm L} \equiv C^S_{RL}\pm C^S_{LL}$ and $C^V_{\pm L} \equiv C^V_{LL}\pm C^V_{RL}$. In deriving these formulas, the authors of \cite{Iguro:2018vqb} use the NLO results of the heavy quark effective theory from \cite{Bernlochner:2017jka} for the hadronic matrix elements. Similar numerical formulas can be found in the literature, e.g.\ \cite{Azatov:2018knx,Angelescu:2018tyl,Asadi:2018sym,Blanke:2018yud}.

As for $\Rjpsi$, there are different calculations for the relevant form factors. In this work we follow the calculation in \cite{Watanabe:2017mip} which, in turn, is based on the form factors calculated in \cite{Wen-Fei:2013uea} using the perturbative QCD factorization. Using these form factors we can calculate the numerical contribution of different WCs to $\Rjpsi$ 
\bea
\label{eq:Nformulasrjpsi}
& \Rjpsi =0.289\Big(0.98|\CVmL|^2+0.02|\CVpL|^2 + 0.05|\CSmL|^2+10.67|\CTLL|^2\\
&\qquad\qquad\qquad + {\rm Re}\left[\CVmL(0.14(\CSmL)^*-5.15(\CTLL)^*) \right]+0.24{\rm Re}\left[\CVpL (\CTLL)^*\right]\Big),\\
\eea
which also indicates that we find $\Rjpsi^{SM}=0.289$, compatible with various other calculations in the literature \cite{Dutta:2017xmj,Watanabe:2017mip,Tran:2018kuv,Cohen:2018dgz,Leljak:2019eyw}. Using other calculations for the form factors would result in different numerical formulas and may affect our final conclusions regarding the maximum attainable value of $\Rjpsi$. This merits further study. However, it is worth noting that our method for maximizing it remains completely general and unchanged and can be adapted to any future version of the numerical formula.

We will be interested in calculating the following quantities:
\beq\label{maxobs}
{\rm max}\,\, F^L_{D^*} \big|_{\RD,\RDs,{\rm Br}(B_c\to\tau\nu)},\qquad {\rm max}\,\, \Rjpsi \big|_{\RD,\RDs,{\rm Br}(B_c\to\tau\nu)}
\eeq
where the global maximum is taken over the full space of WCs with LH neutrinos. (Again, see the end of this section for a generalization to LH+RH neutrinos.) This is a 10 real-dimensional space, making the maximization of $\FLDs$ and $\Rjpsi$ seem like a daunting, if not impossible task. Yet we will accomplish this task by leveraging several properties of the above numerical formulas:
\begin{itemize}
\item All these observables can be written as 
\beq\label{Oquadform}
\CO = z_5^\dagger M_\CO  z_5 = x_5^T M_\CO x_5+y_5^T M_\CO y_5,
\eeq
where 
\begin{eqnarray}
\label{eq:zbardef}
z_5=x_5+iy_5=(\CVmL, \CVpL, \CSmL, \CSpL, \CTLL),
\end{eqnarray}
and the $M_\mathcal{O}$ matrices are real and positive semidefinite.

\item There is one overall rephasing freedom in defining the WCs, i.e. by multiplying all the WCs by a common phase the prediction for these observables does not change.

\end{itemize}

Using these properties (in particular the first one), we can prove that the maxima  (\ref{maxobs}) actually exist. 
We observe that the $M_{\RD}$ and $M_{\RDs}$ matrices in \eqref{Oquadform} have orthogonal null vectors corresponding to $C^S_{-L}$, $\CVmL$ and $C^S_{+L}$, respectively. Hence, fixing $\RD$ and $\RDs$ results in a compact space in the full WC space. Any function on a compact space must have a maximum somewhere in that space. 

We can also prove that the global maximum occurs at real values of the WCs (modulo the overall rephasing invariance). The proof uses the method of Lagrange multipliers. Let's define (for $\CO=\FLDs$ and $\Rjpsi$):
\bea
\tilde \CO &= {\mathcal O} - \lambda_1(\RD-R_D^{(0)}) - \lambda_2(\RDs-R_{D^*}^{(0)})-\lambda_3(\Bc-{\rm Br}(B_c\to\tau\nu)^{(0)}) \cr
 &= x_5^T (M_{\mathcal O}-\lambda_1 M_D-\lambda_2 M_{D^*}-\lambda_3 M_{B_c})x_5 + y_5^T (M_{\mathcal O}-\lambda_1 M_D-\lambda_2 M_{D^*}-\lambda_3 M_{B_c})y_5 \cr
 &\qquad +\lambda_1 R_D^{(0)}+\lambda_2R_{D^*}^{(0)}+\lambda_3{\rm Br}(B_c\to\tau\nu)^{(0)}
\eea
Setting the derivatives of $\tilde \CO$ with respect to $x_5$ and $y_5$ to zero yields 
\beq
\label{eq:xymax}
(M_{\mathcal O}-\lambda_1 M_D-\lambda_2 M_{D^*}-\lambda_3 M_{B_c})x_5=(M_{\mathcal O}-\lambda_1 M_D-\lambda_2 M_{D^*}-\lambda_3 M_{B_c})y_5=0
\eeq
The matrix $M_{\tilde\CO}\equiv M_{\mathcal O}-\lambda_1 M_D-\lambda_2 M_{D^*}-\lambda_3 M_{B_c}$ must be degenerate for this equation to have non-trivial solutions. Yet we cannot tune the $\lambda$s to get more than one zero eigenvalue.\footnote{A proof for generic matrices: in order for $M_{\tilde \CO}$ to be rank less than 4, all of its first minors must be zero. There are 25 such minors, generically independent. So it is impossible to set them all to zero using just three parameters $\lambda_{1,2,3}$. We explicitly check that this argument is true for the matrix combination in \eqref{eq:xymax}.} As a result, the null space is one-dimensional, which means $x_5$ and $y_5$ are parallel to each other. Using the rephasing invariance we can set $y_5=0$, i.e. the WCs at the global maximum can all be taken real.\footnote{As a side note, we can check that the number of unknowns and number of equations match. There are three remaining constraints to satisfy, and three unknowns: $\lambda_2$, $\lambda_3$ and  the modulus of the null vector $x_5$. }

The proof trivially extends to the case of fixing a WC to a particular value. For instance, later we will be interested in fixing $|C^V_{RL}|$ to some value and maximizing the observables with respect to all the other WCs. In that case, we can simply add another quadratic constraint $|C^V_{RL}|^2=(|C^V_{RL}|^2)^{(0)}$ to the mix and the above argument proceeds exactly as before.

So for the rest of the paper we will restrict to real WCs without loss of generality. This reduces the parameter space from $10\to 5$ real dimensional. With the three constraints  $R_D=R_D^0$, $\RDs=R_{D^*}^0$ and ${\rm Br}(B_c\to\tau\nu)=B_c^0$ it amounts to maximizing in 2 real dimensions, or with an additional WC held fixed, in just 1 real dimension. 

Finally, we comment on the generalization to LH+RH neutrinos. Since there is no interference between LH and RH neutrinos, all the numerical formulas in the presence of both types of neutrinos are of the form $z_5^\dagger M z_5+\tilde z_5^\dagger M \tilde z_5$ where $\tilde z_5$ refers to the RH neutrino Wilson coefficients \cite{Asadi:2018sym}. So the Lagrange multiplier argument proceeds as before, and $\tilde z_5$ functions as ``additional imaginary parts", i.e.\ there is an enhanced $SO(4)$ symmetry at the global maximum that allows us to rotate $x_5$, $y_5$, $\tilde x_5$ and $\tilde y_5$ into one another. Thus the global maximum cannot be changed by including RH neutrinos and all of our conclusions derived below which assume only LH neutrinos will be robust.

\section{Maximizing the observables: global maxima}
\label{sec:max}

After we have shown that the maximization problem can be restricted to the real parts of the (LH neutrino) Wilson coefficients without loss of generality, the parameter space is already greatly reduced, and the remaining steps are straightforward if tedious. We perform a series of transformations to the WCs (rotations, shifts and rescalings) so that we can solve the constraints $R_D=R_{D}^0$, $\RDs=R_{D^*}^0$ and ${\rm Br}(B_c\to\tau\nu)=B_c^0$  analytically and simply. This allows the rest of the maximization (over just 2 real dimensions) to be handled numerically. We provide further details on these steps in App.~\ref{sec:appx}. Here we simply present the results.

The results for $\FLDs$ and $\Rjpsi$ are shown in tables \ref{tab:wcsglobalFLD} and \ref{tab:wcsglobalRJpsi} with $\RD$ and $\RDs$ fixed to their world averages and different values of ${\rm Br}(B_c\to\tau\nu)$. We note how similar the numbers are for $\FLDs$ and $\Rjpsi$. It would be interesting to dig deeper into the reasons for this. It is tantalizing and hints at a common NP origin for the two discrepancies. 

Regarding the values of the WCs at the global maxima, there are a few interesting features. In particular, we find a large value of $\CVRL$ and $\CTLL$,\footnote{Notice that all the existing models in the literature generate a tensor WC with association with a scalar WC of $C^S_{LL}\sim 8 \CTLL$ in the IR; hence, having $\CTLL \sim 0.3$ in the IR implies scalar WCs of around $2.4$.} and a substantial cancellation of the SM contribution to $\CVLL$. These are in fact generic features we find in the combination of the WCs that maximize $\FLDs$ and $\Rjpsi$ for other values of $\RDRDs$ and $\Bc$ as well. This suggests that any NP origin of $\FLDs$ and $\Rjpsi$ may be nonminimal, in order to give rise to all of these WCs.

\begin{table}
\resizebox{1.\columnwidth}{!}{
\begin{tabular}{|c|c|c|c|c|c|c|c|c|c|}
\hline 
$\CSRL$ & $\CSLL$ & $\CVLL$ & $\CVRL$ & $\CTLL$ & $\RD$ & $\RDs$ & $\FLDs$ & $\Rjpsi$ & $\Bc$\\ 
\hline 
-0.669 & -0.884 & 0.097 & 2.029 & -0.329 & 0.407 & 0.304 & 0.620 & 0.406 & 0.023 \\
\hline 
-0.791&-0.739&0.118&1.977&-0.302 & 0.407 & 0.304 & 0.638 & 0.410 & 0.1 \\
\hline 
-0.972&-0.555&0.142&1.948&-0.298 & 0.407 & 0.304 & 0.662 & 0.412 & 0.3 \\
\hline 
\end{tabular} 
}
\caption{The combination of WCs that maximize $\FLDs$ for the global average of $\RDRDs$ and with various values of $\Bc$. All these combinations exhibit a large value of $\CVRL$ and $\CTLL$; the SM contribution of $\CVLL=1$ is also largely canceled. }
\label{tab:wcsglobalFLD}
\end{table}

\begin{table}
\resizebox{1.\columnwidth}{!}{
\begin{tabular}{|c|c|c|c|c|c|c|c|c|c|}
\hline 
$\CSRL$ & $\CSLL$ & $\CVLL$ & $\CVRL$ & $\CTLL$ & $\RD$ & $\RDs$ & $\FLDs$ & $\Rjpsi$ & $\Bc$\\ 
\hline 
-0.659 & -0.857 & 0.109 & 1.967 & -0.286 & 0.407 & 0.304 & 0.620 & 0.409 & 0.023 \\
\hline 
-0.787&-0.726&0.124&1.948&-0.282 & 0.407 & 0.304 & 0.637 & 0.410 & 0.1 \\
\hline 
-0.967&-0.542&0.147&1.919&-0.277 & 0.407 & 0.304 & 0.660 & 0.413 & 0.3 \\
\hline 
\end{tabular} 
}
\caption{The combination of WCs that maximize  $\Rjpsi$ for the global average of $\RDRDs$ and with various values of $\Bc$. Intriguingly, the WCs at the global maximum of $\Rjpsi$ exhibit very similar features to those at the global maximum of $\FLDs$.}
\label{tab:wcsglobalRJpsi}
\end{table}

\begin{figure}
\includegraphics[scale=0.8]{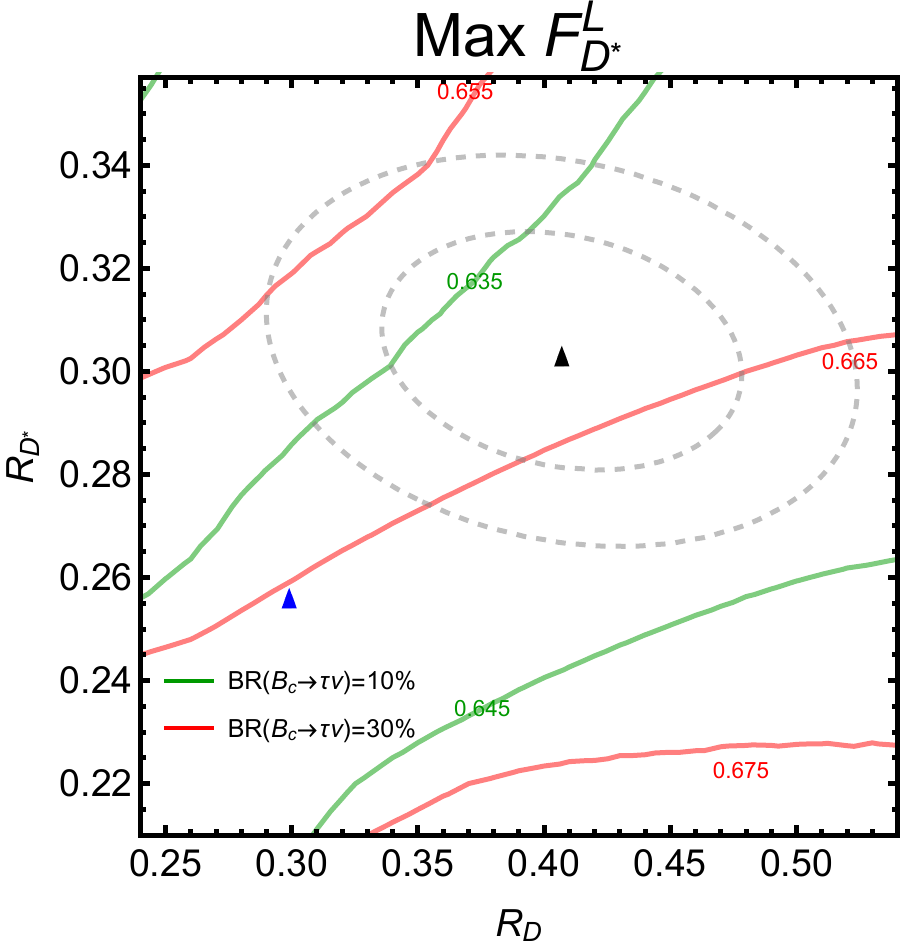}
\includegraphics[scale=0.8]{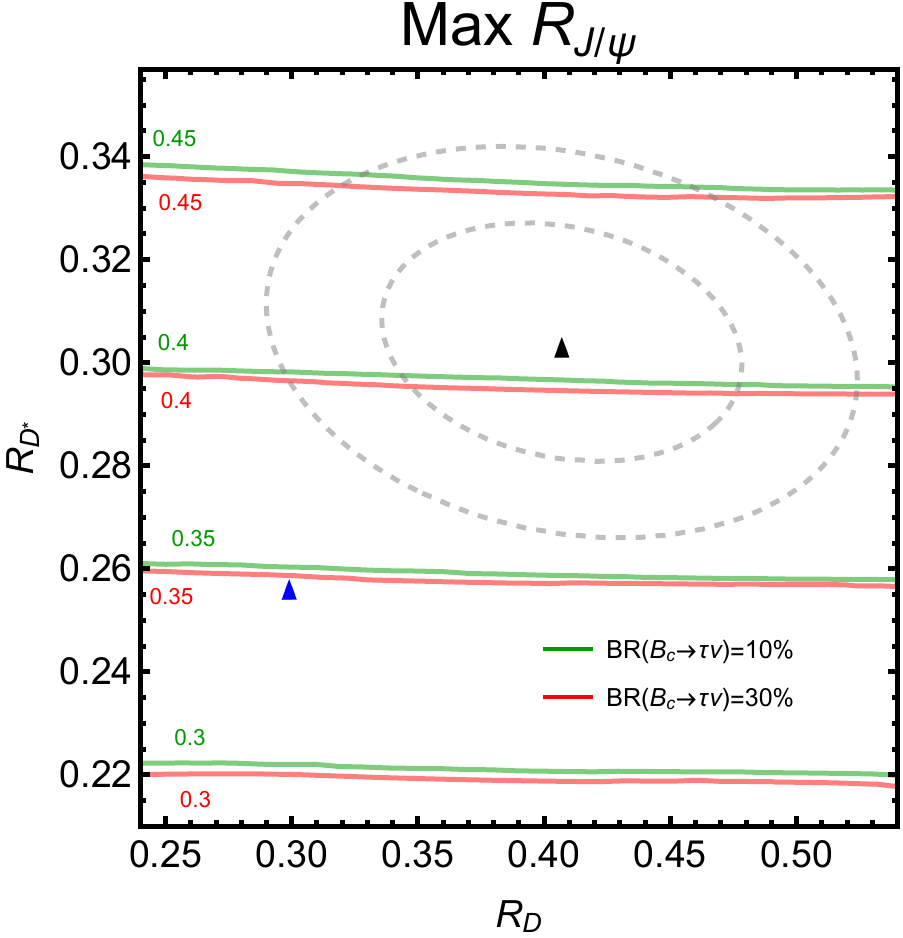}
\caption{The maximum attainable $\FLDs$ (left) and the maximum attainable $\Rjpsi$ (right) for different values of $\Bc$ and $\RDRDs$. The green and red contours correspond to $\Bc=10\%$ and $\Bc=30\%$, respectively. The blue (black) triangle indicates the SM predictions (the world-averaged measured values) of $\RDRDs$ while the dashed gray ellipses are contours of $1$ and $2\sigma$ around the world-average measured values. These figures indicate that indeed there exists a combination of the WCs that can explain the observed value of $\FLDs$ from \eqref{eq:fldexp}; yet, there are no combinations of these WCs that can reach the $1\sigma$ range of the observed $\Rjpsi$ value in \eqref{eq:rjpsiexp}. }
\label{fig:global}
\end{figure}

In Fig.~\ref{fig:global}, we find the maximum of $\FLDs$ or $\Rjpsi$ over all the WCs for different values of $\Bc$ and $\RDRDs$. 
The figures indeed show the observed $\Rjpsi$ is not obtainable anywhere in the parameter space of the most general dimension 6 effective Hamiltonian with LH and RH neutrinos. If the future measurement of $\Rjpsi$ remains at its present value, then it will be a very sharp contradiction with the present framework. It could point at either a significant revision to the hadronic form factors for $\Rjpsi$, or to NP that is somehow not captured by the dimension 6 effective Hamiltonian (for instance, very light mediators).

Meanwhile, we see that the observed value of $\FLDs$ is attainable everywhere in the 1 or 2$\sigma$ ellipse of the measured world average $\RD$, $\RDs$. However, no known models currently can give rise to such a large value of $\FLDs$ \cite{Iguro:2018vqb,Blanke:2018yud}. 
 This could be due to the fact that we seem to need a combination of all the WCs to have a large enhancement to $\FLDs$, as suggested by Tab.~\ref{tab:wcsglobalFLD}, which can not be achieved with any of the existing minimal models. It could also be due to the fact that enhanced $\FLDs$ seems to require a large value of $C^V_{RL}$, which is well-known to be challenging. We will discuss $C^V_{RL}$ further in the next section.

\section{Maximizing the observables: holding WCs fixed}
\label{sec:resultfixed}

We can also treat any of the WCs as a constant and go through a similar series of transformations as above, in order to maximize $\FLDs$ and $\Rjpsi$ when holding that WC fixed. This allows us to study that WC's contribution to $\FLDs$ and $\Rjpsi$ in further detail.

\begin{figure*}
\includegraphics[scale=0.7]{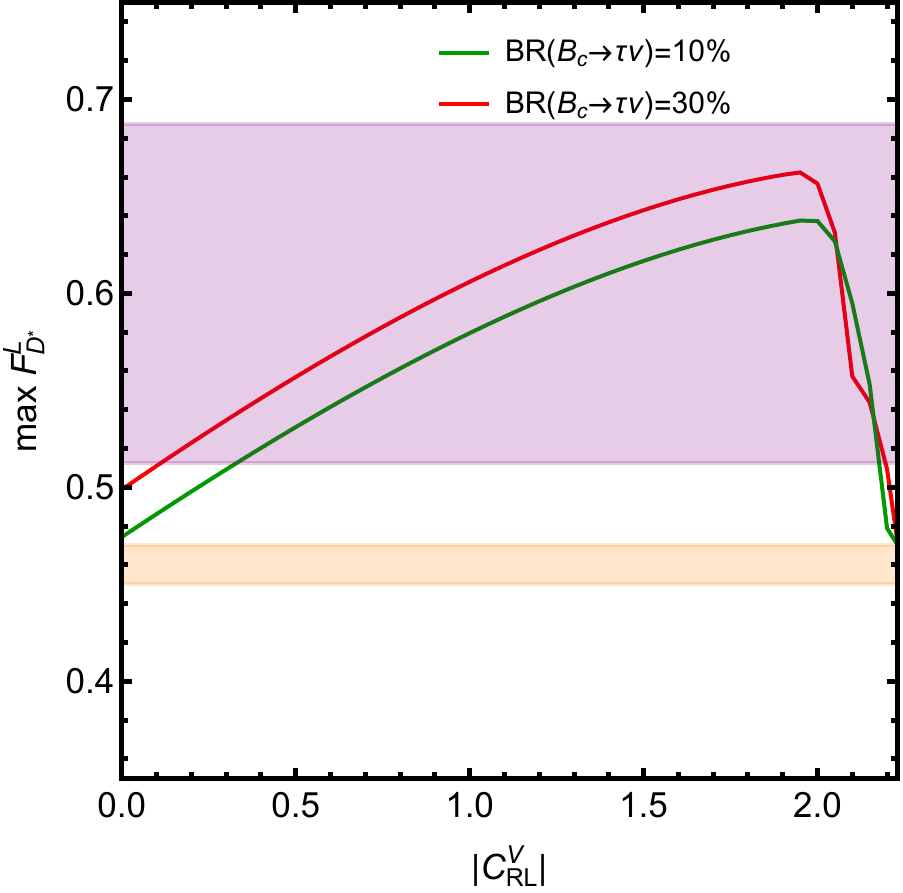}
\includegraphics[scale=0.7]{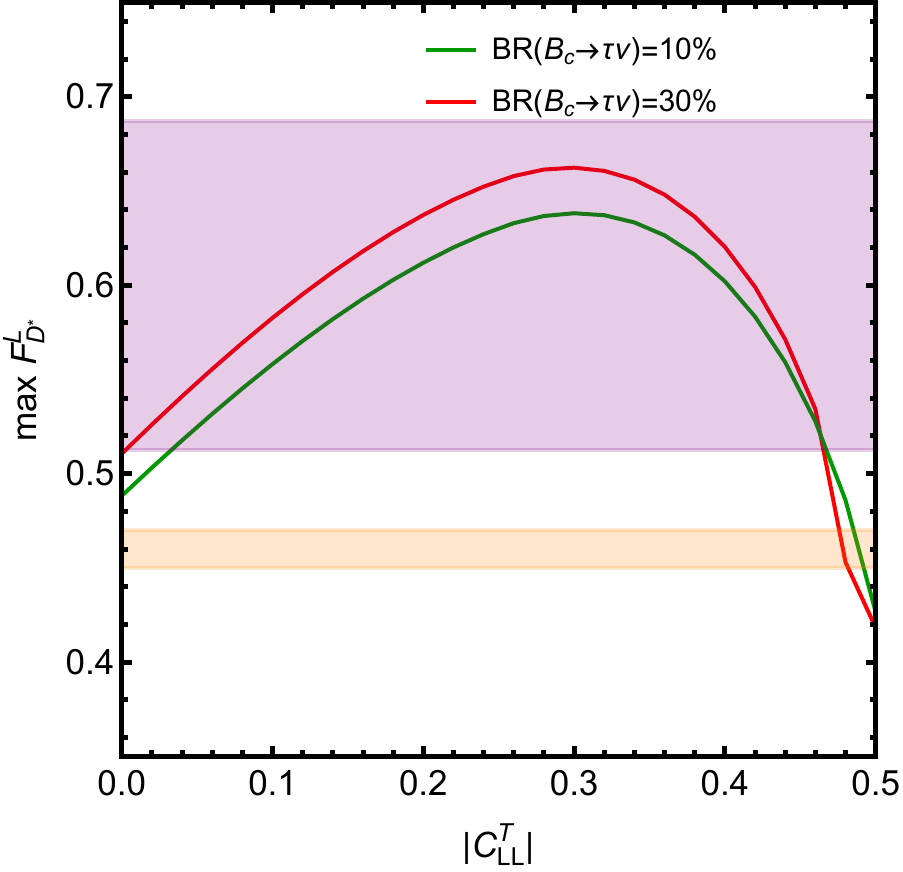}
\includegraphics[scale=0.7]{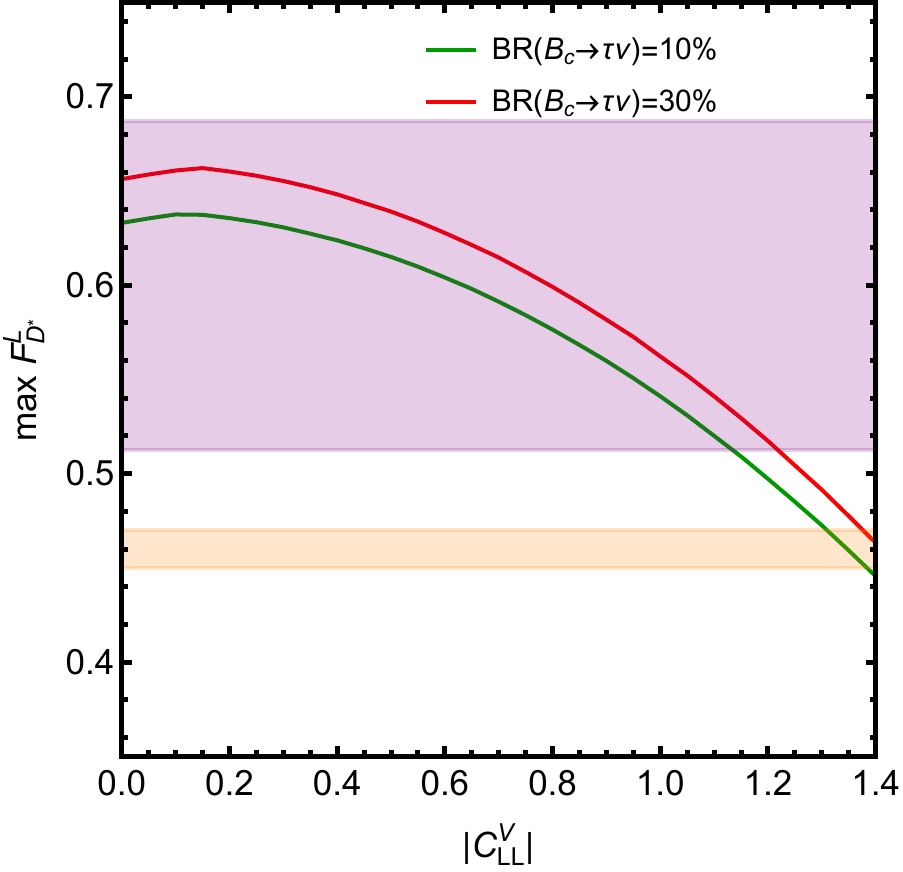}
\caption{	The maximum attainable $\FLDs$ as a function of WCs $\CTLL$, $\CVRL$, or $\CVLL$; in each plot we marginalize over other WCs, given the constraints $\RD=0.4$ and $\RDs=0.3$. The green and red curves correspond to $\Bc=10\%$ and $\Bc=30\%$, respectively. The purple (orange) band shows the $1\sigma$ error bar around the central observed value (SM prediction) of $\FLDs$. These figures highlight the necessity of NP with all of these WCs in order to explain the observed $\FLDs$.}
\label{fig:fld_cvct}
\end{figure*}

Going through the procedure above for all different WCs we find interesting results for the contributions of $\CTLL$, $\CVLL$, and $\CVRL$ to $\FLDs$. In Fig.~\ref{fig:fld_cvct} we show the maximum attainable value of $\FLDs$ as a function of these three WCs, and in Tab.~\ref{tab:wcs} we report a few benchmark points maximizing $\FLDs$ for a fixed $\CVRL$. 
These  clearly suggest that in order to explain the observed $\FLDs$ in \eqref{eq:fldexp}, we need non-zero values for all of these WCs from NP. In Fig.~\ref{fig:fld_cvct}, if we go to larger values of the fixed WC in each plot, it becomes impossible to satisfy the constraints on $\RDRDs$. 

Most notably, Fig.~\ref{fig:fld_cvct} demonstrates that in order to explain the observed $\FLDs$ from \eqref{eq:fldexp}, NP should give rise to sizable $\CVRL$. There are currently no models in the literature generating this WC. In fact, there are strong general arguments against its existence. It violates $SU(2)_L$ and $U(1)_Y$ so it must be higher effective dimension (at least dimension 8).\footnote{As discussed in \cite{Alonso:2015sja,Cata:2015lta}, one can generate this operator at dimension 6 in SMEFT but only by integrating out an off-shell $W$; since the couplings of the $W$ to the leptonic side are flavor-universal, this can not explain our anomalies, which require some LFU violation.}

\begin{table}
\resizebox{1.\columnwidth}{!}{
\begin{tabular}{|c|c|c|c|c|c|c|c|c|c|}
\hline 
$\CSRL$ & $\CSLL$ & $\CVLL$ & $\CVRL$ & $\CTLL$ & $\RD$ & $\RDs$ & $\FLDs$ & $\Rjpsi$ & $\Bc$\\ 
\hline 
0.330 & 0.152 & 1.012 & -0.3 & 0.092 & 0.400 & 0.300 & 0.510 & 0.340 & 0.1 \\
\hline 
0.481 & 0.321 & 0.890 & -0.5 & 0.118 & 0.400 & 0.300 & 0.532 & 0.347 & 0.1 \\
\hline 
0.614 & 0.471 & 0.764 & -0.7 & 0.143 & 0.400 & 0.300 & 0.552 & 0.355 & 0.1 \\
\hline 
0.785 & 0.665 & 0.567 & -1 & 0.180 & 0.400 & 0.300 & 0.580 & 0.365 & 0.1 \\
\hline 
\end{tabular} 
}
\caption{Benchmark points that can reach the maximum $\FLDs$ with a particular $\CVRL$ and fixed $\RDRDs$ and $\Bc$. The $\Rjpsi$ with the same set of WCs is calculated as well; these values of $\Rjpsi$ are very close to the maximum attainable $\Rjpsi$ with the same $\CVRL$, see fig.~\ref{fig:rjpsi_cvct}. }
\label{tab:wcs}
\end{table}

\begin{figure*}
\includegraphics[scale=0.7]{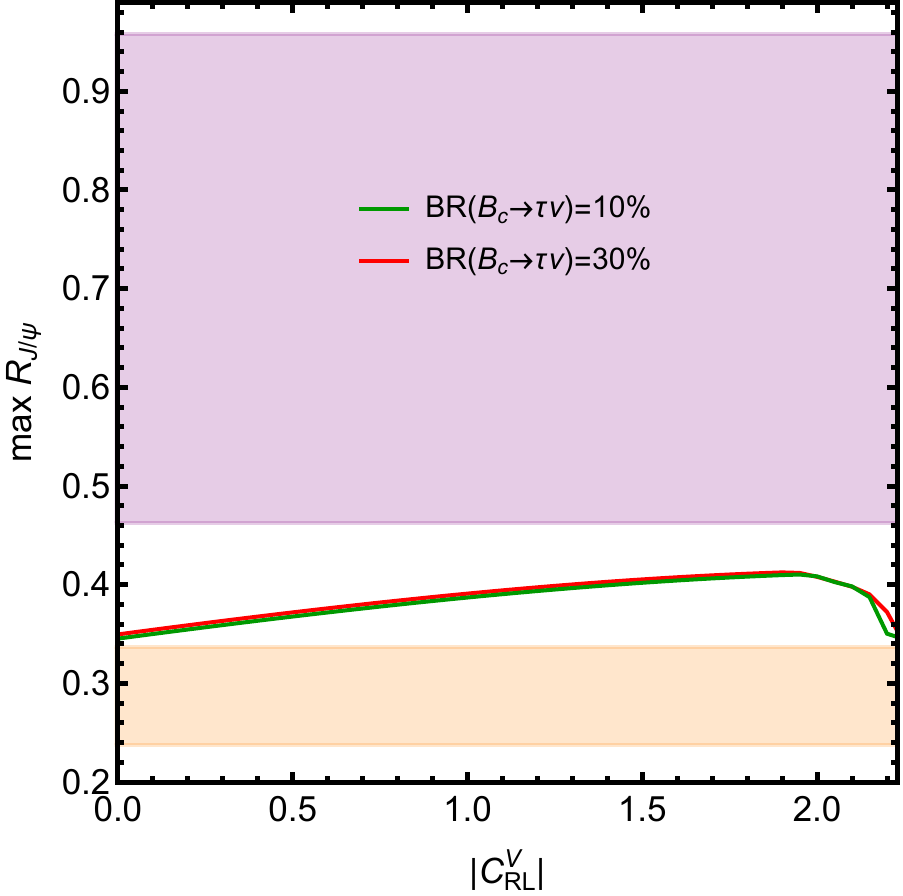}
\includegraphics[scale=0.7]{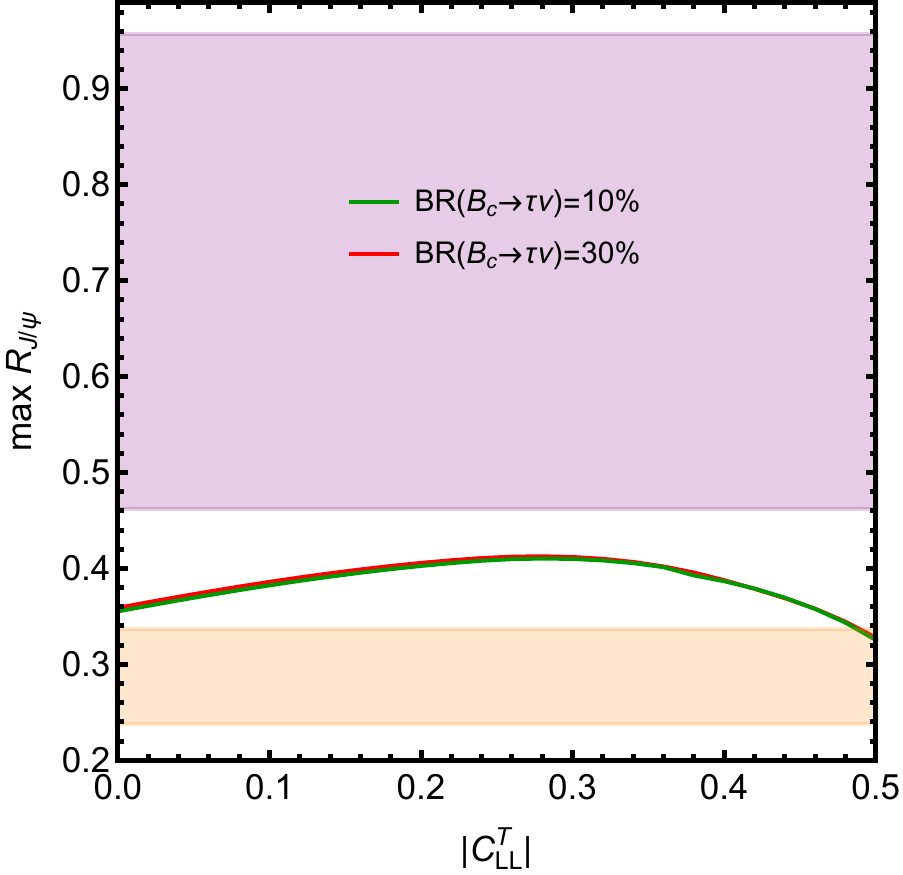}
\includegraphics[scale=0.7]{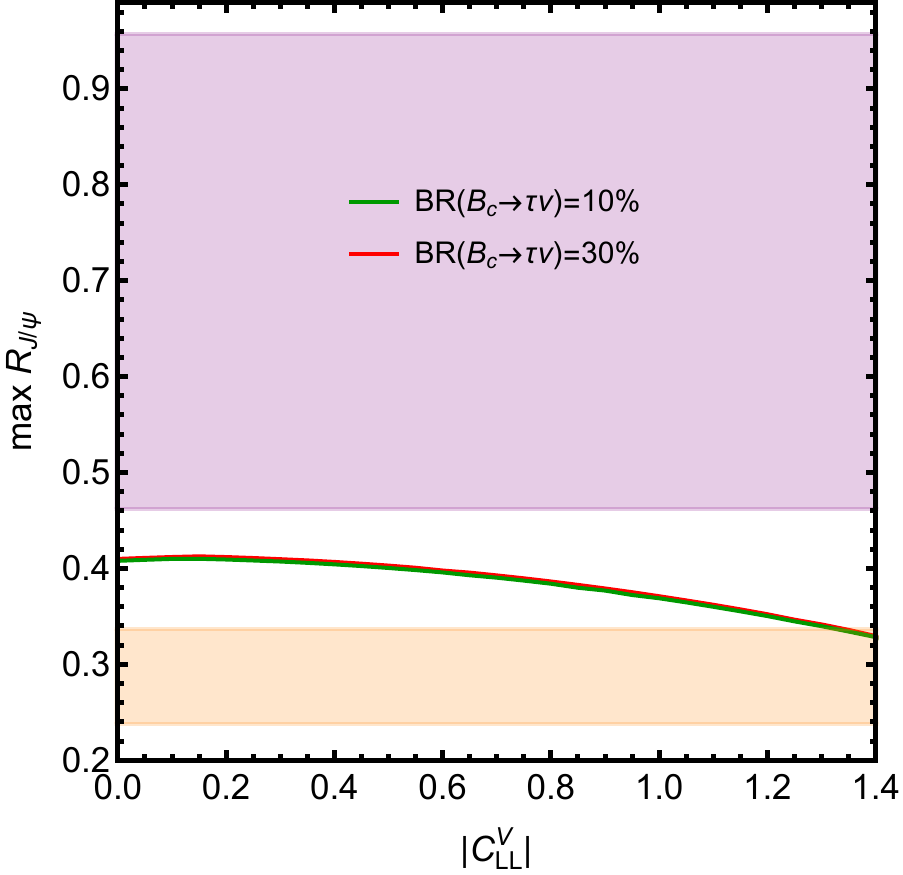}
\caption{The maximum attainable $\Rjpsi$ as a function of WCs $\CTLL$, $\CVRL$, or $\CVLL$; in each plot we marginalize over other WCs. The colors and bands are as in fig.~\ref{fig:fld_cvct}. We see that we can not even reach the $1\sigma$ range of the observed $\Rjpsi$ for any values of the WCs.}
\label{fig:rjpsi_cvct}
\end{figure*}

As we saw in Fig.~\ref{fig:global}, there is no point in the parameter space of the dimension 6 effective Hamiltonian consistent with the measured values of $\RD$ and $\RDs$ that can explain the observed value of $\Rjpsi$. For completeness, we elaborate on this by studying the effect of each individual operator on $\Rjpsi$. The maximum $\Rjpsi$ attainable with fixed values of certain WCs is depicted in Fig.~\ref{fig:rjpsi_cvct}. 
We further include the prediction for $\Rjpsi$ with the WCs in Tab.~\ref{tab:wcs} that maximize $\FLDs$ for any given $\CVRL$; these benchmark points can almost reach the maximum attainable $\Rjpsi$ as well.

 \vskip0.5cm

\noindent\textbf{Note added.} During the final stages of this work \cite{Murgui:2019czp} appeared on arXiv with partially overlapping results concerning $\FLDs$ and $\Rjpsi$. The authors of \cite{Murgui:2019czp} carried out an extensive global fit of various observables with the effective operators involving LH neutrinos and arrived at a similar conclusion as in this work regarding the importance of $\CVRL$ in explaining $\FLDs$.

\section*{Acknowledgments}
 We thank Marat Freytsis and Ryoutaro Watanabe for helpful discussions. This work is supported by DOE grant DOE-SC0010008.

\appendix

\section{Details on maximizing the observables}
\label{sec:appx}

We now provide some details to our procedure. We hope these details will prove useful to others who may be interested in maximizing other observables in the future (or replicating our analysis). 

The first step is to solve the equation of $\Bc$ for $\CSmL$,
\begin{equation}
\CSmL = \frac{1}{4.33} \left(	 e^{i\xi} \mathcal{R}_{B_c}	 - \CVmL	\right),
\label{eq:bcsolved}
\end{equation}
where $\xi$ is an arbitrary phase and we have defined
\begin{equation}
\mathcal{R}_{B_c} \equiv \sqrt{\frac{\Bc}{\Bc^{SM}}}.
\label{eq:RBc}
\end{equation}
We can use the phase invariance mentioned earlier to fix the value of $\xi$ to any number in order to simplify the calculation; in our analysis, we use $\xi=\pi$. With this choice of $\xi$ we explicitly break the symmetry between the contribution of real and imaginary parts of the WCs to various observables and exhaust the freedom in rephasing the WCs. 

Next, we perform the following transformation (which is a combination of rotations, shifts and rescalings) on the WCs: 
\begin{eqnarray}
\label{eq:finalnumericsLHnoCVRLmtx}
\left(	\begin{matrix}
\CSpL \\ \CVpL \\\CVmL \\ \CTLL
\end{matrix}	\right)	& = & \left(	\begin{matrix}
1.8108 &3.7863  &-2.1150 &0 \\
0 &-5.1839  & 2.8958 &0 \\
0 &13.3846 &-0.4787 & -1  \\
0 & 4.2232&  -0.1510 & 0
\end{matrix}	\right) \left(	\begin{matrix}
\tilde C^S_{+L} \\ \tilde C^V_{+L}  \\ \tilde C^V_{-L} + 0.0114  \mathcal{R}_{B_c}  \\ \tilde C^T_{LL}
\end{matrix}	\right), 
\end{eqnarray}
in order to simultaneously diagonalize the quadratic terms in $\RD$ and $\RDs$:
\bea
\label{eq:finalnumericsLHnoCVRL}
& \RD=(\tilde C^S_{+L})^2 + \tilde x_3^T M_{D} \tilde x_3 \\
& \RDs= \tilde x_3^T \tilde M_{D^*}\tilde x_3 + v_{D^*}^T \tilde x_3 + A_{D^*}\\
\eea
Here $\tilde x_3\equiv(\tilde C^V_{+L},\,\,\,\tilde C^V_{-L},\,\,\, \tilde C^T_{LL})^T$ and  
\begin{eqnarray}
\label{eq:afterrotations}
&\small{\tilde{M}_{D} = \left( \begin{matrix}
1 &0&0\\
0&1&0 \\
0&0&0 \\
\end{matrix} \right),
\tilde{M}_{D^*} = \left( \begin{matrix}
26.7838 &0&0\\
0&0.0553&0 \\
0&0&0.2388 \\
\end{matrix} \right),
 v_{D^*}=\left( \begin{matrix}
-0.0727\mathcal{R}_{B_c} \\
0.0026\mathcal{R}_{B_c} \\
0 \\
\end{matrix} \right),
 A_{D^*}=0.0005 \mathcal{R}_{B_c}^2.\nonumber
}\\
\end{eqnarray}

Under this transformation, the observables become:
\bea
& \RDs\FLDs = \tilde x_3^T \tilde M_{F} \tilde x_3 + v_{F}^T\tilde x_3+A_F\\
& \Rjpsi = \tilde x_3^T \tilde M_{J/\psi} \tilde x_3 + v_{J/\psi}^T\tilde x_3+A_{J/\psi}\\
\eea
where
\bea
\tilde{M}_{F}=\left( \begin{matrix}
5.6079 &-0.2005&-0.4042\\ 
-0.2005&0.0072&0.0145 \\
-0.4042&0.0145 &0.1105 \\
\end{matrix} \right),
\quad v_F=\left( \begin{matrix}
-0.0639\mathcal{R}_{B_c} \\
0.0023\mathcal{R}_{B_c} \\
0.0029\mathcal{R}_{B_c} \\
\end{matrix} \right),
\quad A_F=0.0004\mathcal{R}_{B_c}^2, \\
\tilde{M}_{J/\psi}=\left( \begin{matrix}
18.8505 &-0.3420&-0.5463\\
-0.3420&0.0368&0.0195 \\
-0.5463&0.0195&0.2756\\
\end{matrix} \right),
\quad v_{J/\psi}=\left( \begin{matrix}
-0.0945\mathcal{R}_{B_c} \\
0.0034\mathcal{R}_{B_c} \\
0.0017\mathcal{R}_{B_c} \\
\end{matrix} \right),
\quad A_{J/\psi} = 0.0007\mathcal{R}_{B_c}^2 .
\eea
We can go to spherical coordinates in $(\tilde C^S_{+L},\tilde C^V_{+L},\tilde C^V_{-L})$ and solve the $R_D$ constraint for the radial coordinate. Then we can solve the $\RDs$ constraint for $\tilde C^T_{LL}$ which only appears as $(\tilde C^T_{LL})^2$. This leaves behind two angles which we can then easily numerically maximize over and verify explicitly with a plot.

\bibliographystyle{utphys_modified}
\bibliography{bib}

\end{document}